\documentclass[pre,a4paper,oneside,twocolumn,nofootinbib]{revtex4}
\usepackage[utf8x]{inputenc}
\usepackage{latexsym}
\usepackage{fancyhdr}
\usepackage{color} 
\usepackage{soul} 
\usepackage{graphicx}  

\usepackage[dvipsnames]{xcolor}

\usepackage[pdfborder={0 0 0}]{hyperref}

\usepackage[normalem]{ulem} 

\begin{document}

\title{A Gentle Introduction to Scaling Laws in Biological Systems}


\author{Fabiano L. Ribeiro\textsuperscript{1}\textsuperscript{*} and 
	William R. L. S. Pereira\textsuperscript{2}\textsuperscript{$\diamond$}		
		}

\begin{abstract}
This paper investigates the role of size in biological organisms.  More specifically, how the energy demand, expressed by the metabolic rate, changes according to the mass of an organism.  Empirical evidence suggests a power-law relation between mass and metabolic rate, namely allometric law.  For vascular organisms, the exponent $\beta$ of this power-law is smaller than one, which implies scaling economy; that is, the greater the organism is, the lesser energy per cell it demands.  However, the numerical value of this exponent is a theme of an extensive debate and a central issue in comparative physiology.
It is presented in this work some empirical data and a detailed discussion about the most successful theories to explain these issues.
A historical perspective is also shown,  beginning with the first empirical insights in the sec. 19 about scaling properties in biology, passing through the two more important theories that explain the scaling properties quantitatively.  
Firstly, the Rubner model,  that consider organism surface area and heat dissipation to derive $\beta = 2/3$.
Secondly, the West-Brown-Enquist theory,  that explains such scaling properties as a consequence of the hierarchical and fractal nutrient distribution network, deriving $\beta = 3/4$.	
\end{abstract}

\maketitle

\textbf{1} Departmento de Física (DFI), Universidade Federal de Lavras (UFLA), Lavras MG, Brazil; 

\textbf{2} Independent Researcher.


\textbf{*} fribeiro@ufla.br

\textbf{$\diamond$} william.roberto.luiz@gmail.com

%

\section{Introduction}

When we hold a small hamster in our hands, we can feel its fast heartbeat, with approximately 450 bpm.  
In turn, if we listen to the heartbeat of an elephant, we will realize that the heart rate of this large mammal is extremely slow, approximately 30 bpm. In fact, this higher heart rate of hamsters reveals a higher \textit{metabolic rate} of the rodent compared to that of the elephant. But why is it greater? What does this mean and why does it happen?

\begin{figure}
	\begin{center}
		\includegraphics[width=\columnwidth]{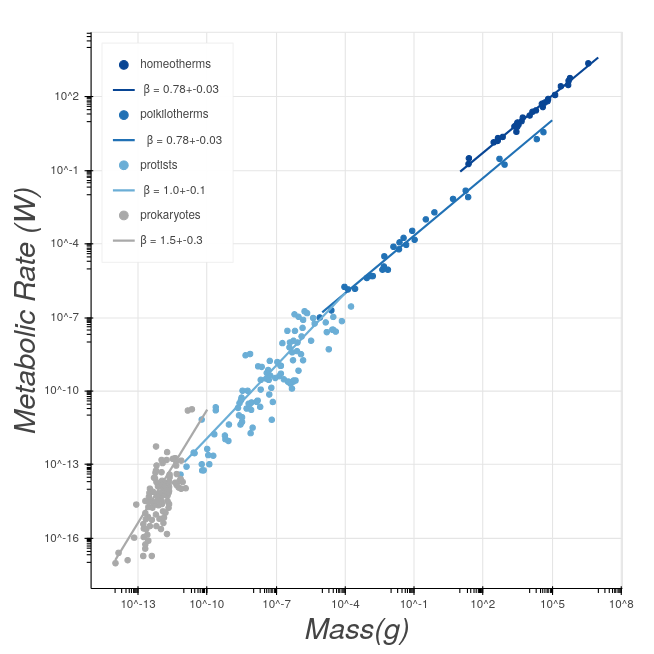}
	\end{center}
	\caption{ \label{Fig_lei34} 
		Metabolic rate as a function of body mass in different taxa, from unicellular beings ($10^{-14}$ grams) to the largest mammals ($10^{8}$ grams)  (see also fig.~(\ref{Fig_ani-massa})). A valid power law of the type
		$B = B_0 M^{\beta}$ (straight line captures the trend of points) is observed for all taxonomic groups. The parameter $B_0$ varies from group to group, and $\beta$  is approximately constant and sublinear ($\beta < 1$) in beings with a mass of approximately $10^{-5}$g or higher. Protists have linear behaviour ($\beta \approx 1$), and bacteria have a superlinear behaviour ($\beta > 1$). The data were extracted directly from the references \cite{Allman1999}(homeotherms), \cite{Hemmingsen1960}(homeotherms and poikilotherms), and \cite{shiftspnas2010}(prokaryotes and protists). 
	}
\end{figure}

\begin{figure}
	\begin{center}
		\includegraphics[width=\columnwidth]{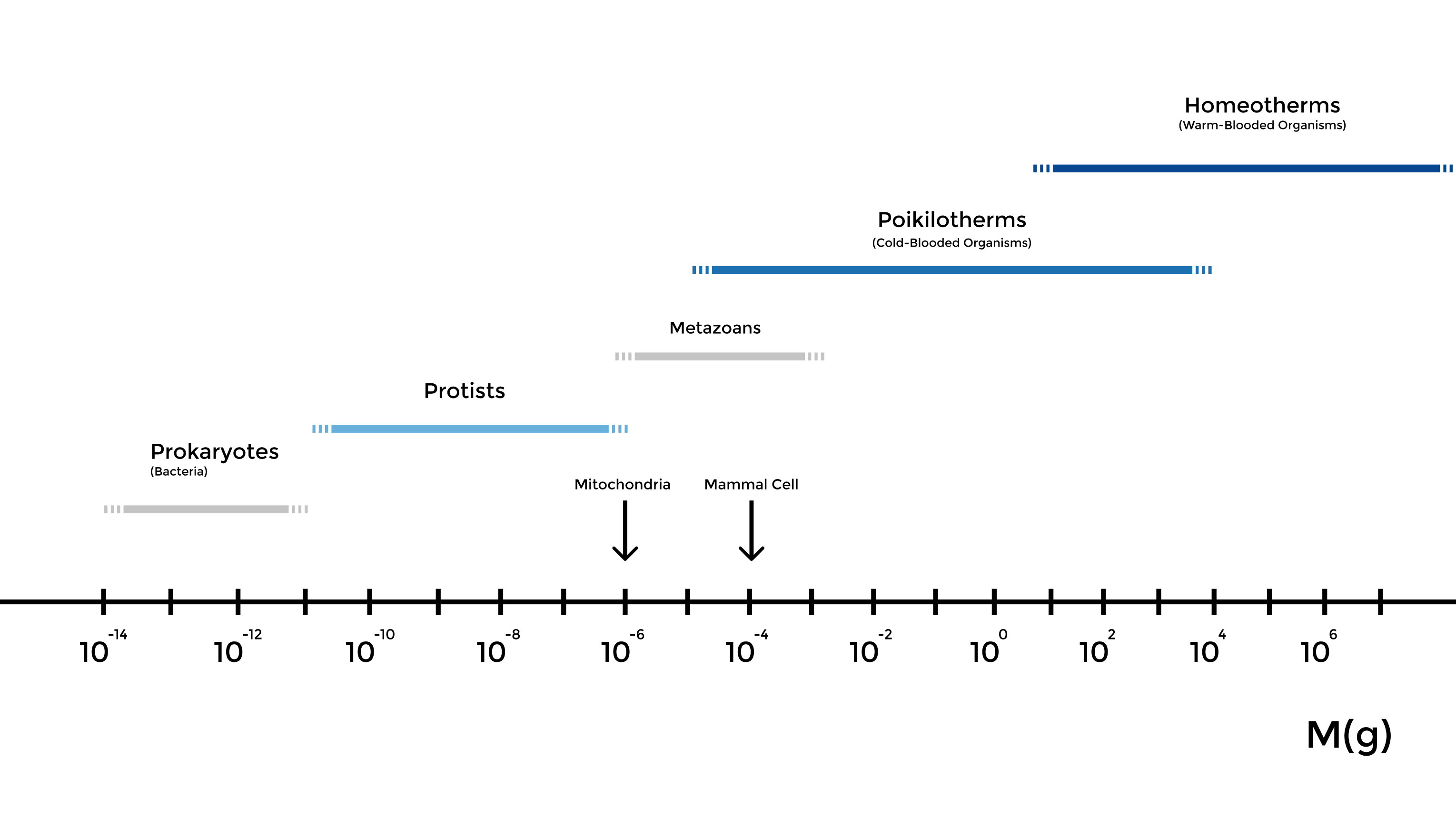}
	\end{center}
	\caption{ \label{Fig_ani-massa} 
		Biological systems covers around 22 orders of magnitude, from unicellular beings to the largest mammals.}
\end{figure}

The  \textit{metabolic rate} is the mean value of energy, per unit of time, used by an organism to perform its vital functions. This energy is obtained from food, water, air, light, etc. Figure~(\ref{Fig_lei34}) shows the log–log scale graph of the empirical value of the metabolic rate $B$ (in watts) as a function of mass $M$ (in grams) of organisms of different taxonomic groups. The straight lines in this figure show that all groups can be described individually by a power law of the type

\begin{equation}\label{eq_halome}
B = B_0 M^{\beta}.
\end{equation}
This equation was called the \textit{allometric equation} by Huxley in 1932 \cite{HUXLEY1932}, where $B_0$ is the \textit{allometric constant} and $\beta$  is the \textit{allometric exponent}. The straight lines in Figure~(\ref{Fig_lei34})  show a compatibility of the data with the allometric law. This law covers 22 orders of magnitude (see fig.~(\ref{Fig_ani-massa})), from unicellular beings ($10^{−14}$ grams) to the largest mammals ($10^8$ grams), and it differs only in the values of the parameters for certain taxonomic groups. The data present fig.~(\ref{Fig_lei34}) also 
show three different regimes expressed by the scaling exponent: superlinear  ($\beta >1$), linear ($\beta =1$)  and sublinear  ($\beta<1$).

These facts show us that it does not matter how complex a species is, such as in its physiology, its anatomy, or the different types of environments in which it lives. In the end, all animals obey the same rule relating metabolic rate with mass. In the present study, we are interested in investigating the scaling laws behind this allometric equation. More specifically, we will investigate the possible $\beta$ values in different types of animals, in addition to offering explanations for these values.


The allometric equation was first perceived in 1839 by Saurus and Rameaux \cite{Robiquet1839}. These researchers noticed this relationship when they realized that the metabolic rate per weight decreases with increasing animal size. Following \textit{Fourier’s law}\cite{Roselli2011,Ballesteros2018b}, they proposed that the metabolic rate should depend on heat dissipation by the organism. Thus, the numerical value of the allometric exponent would be a natural response to the release of heat by the organism and would make the relationship between the surface area and volume of the organism valid. This idea led to a theoretical exponent  $\beta = 2/3$. Further details regarding this deduction will be presented in section~(\ref{sec_calor}).

At the end of the 19-th century, some experiments were performed to verify the empirical value of the allometric exponent. For example, Rubner  \cite{Rubner1883,bertalanffy1957}  studied dogs and in 1883 found that their energy production per square metre of body surface is constant with the size of the animal, which was evidence in favour of $\beta = 2/3$. More careful experiments were performed at the beginning of the 20th century. Among these studies, we highlight the works of Krogh\cite{Krogh1916}  and that of Kleiber in 1932 (the best known study) \cite{Kleiber1932}. From the data set analysed, an experimental value of $\beta_{exp} \approx 3/4$ was observed, which differed from the theoretical result that was accepted until then.
	It is just the controversy about the scaling exponent different empirical values and theories that this work focuses on.

The work is organized as follows. In the section~(\ref{sec_diff_beta}), some empirical evidence from the literature is presented, showing that the empirical value of the scaling exponent is wildly inaccurate. This section also presents some groups of theories proposed to explain such scaling properties.  The section~(\ref{sec_calor}) presents a model that explains the scaling properties as a consequence of heat dissipation and organism surface area, namely Rubner model, deriving the exponent $ \beta = 2/3 $. In the section~(\ref{secao_rede_distribuicao}), the theory developed by West-Brown-Equisque based on the nutrient distribution network is presented. This theory leads to an exponent $ \beta = 3/4$. The conclusion is presented in section~(\ref{sec_conclusion}).

\section{Different values of $\beta$} 
	\label{sec_diff_beta}
	
\begin{figure}
	\begin{center}
		\includegraphics[width=\columnwidth]{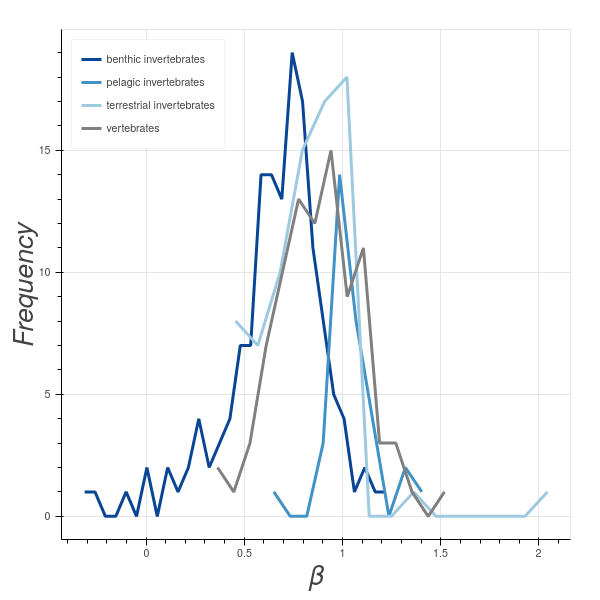}
		\end{center}
	\caption{ \label{fig_beta_distr} 		
		Histograms referring to the number of species during ontogeny (development) and the allometric exponent value for different taxonomic groups. The histograms show the sublinearity of $\beta$ within these taxonomic groups but also show uncertainty about its exact value. 
		The data were taken directly from \cite{Moses2008,Glazier2005b}.
		}
\end{figure}

Currently, we know that organisms larger than  $10^{-5}$g have sublinear regimes (see Figures~(\ref{Fig_lei34})  and   ~(\ref{fig_beta_distr})), with some taxonomic groups best described by $\beta = 2/3$, while the vast majority are better described by  $\beta = 3/4$. Figure~(\ref{fig_beta_distr}), which shows the distribution of $\beta$ values for different taxonomic groups with sufficiently large masses, shows the sublinearity of this exponent among the analysed groups, but it also leaves doubts about the exact value of this exponent.

The theoretical and experimental values of  $\beta$ are a central issue in comparative physiology\cite{Ballesteros2018b}. Only to cite some of the many examples in the literature, Dodds et al. \cite{dodds2001} show that rats smaller than 10 kg are better described by an exponent of $2/3$, while for the other sizes, the exponent of $3/4$ is better. Exponent $2/3$ well describes fish and some types of invertebrates, such as crustaceans and mussels \cite{bertalanffy1957}; exponent $3/4$ fits better \textit{endothermic animals}\footnote{Those that keep a constant internal temperature.} 
(birds and mammals) and reptiles. 
In contrast, organisms with sizes between $10^{−10}$g and $10^{−5}$g, such as \textit{protists}, exhibit a linear behaviour of the allometric law  \cite{shiftspnas2010}. Some types of insects also obey this linear regime \cite{bertalanffy1957}. Finally, we have organisms smaller than $10^{−5}g$, namely, bacteria, which exhibit superlinear behaviour (see Figure~(\ref{Fig_lei34})).



But what mechanisms lead to this scaling law, and why can different $\beta$ values occur? We can think of two types of explanatory approaches. The \textit{ cellular hypothesis} suggests that the allometric law is the result of distinct cellular properties in animals of different sizes. The \textit{scaling hypothesis}, on the other hand,  suggests that this law would be the result of regulatory factors of the organism as a whole.

\begin{figure}
	\begin{center}
		\includegraphics[width=\columnwidth]{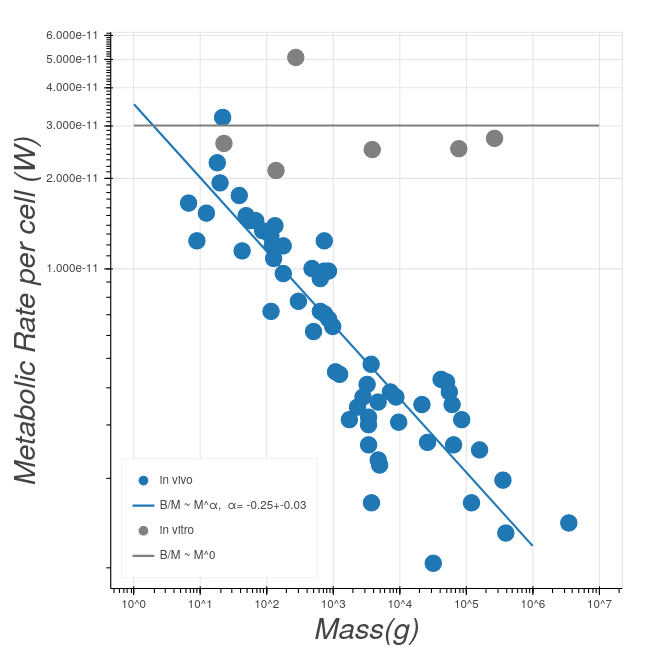}
	\end{center}
	\caption{ \label{Fig_west} 
		Metabolic rate per cell ($B/M$) as a function of the organism mass 
		($M$), analyzed in two contexts: \textit {in vitro} and \textit {in vivo}. 	In  \textit{in vivo}  cells,  the metabolic rate per cell decays with the mass of the organism obeying a power law with exponent $-1/4$. 		
		In contrast,  \textit{in vitro} 
		cells do not show any scale relation, i.e. $(B/M)\sim M^0$  (horizontal line). Data was extracted directly from \cite{West2002a}. }
\end{figure}

Some empirical observations favour the scaling hypothesis. For example, in the experiment described in~\cite{Brown2007}, cells from 10 different mammals were cultured  \textit{in vitro}, i.e., outside their original organisms. This study found that all cells, regardless of the animal of origin, had the same metabolic rate. However, note that the allometric equation~(\ref{eq_halome}) tells us that the metabolic rate per cell $ B/N$, where $N$ is the number of cells of the organism, decreases with its  size  (because 
$M \sim N$, and then  $ B/N \sim B/M \sim M^{\beta -1}$, with $\beta < 1$). 
That is, when \textit{in vivo}, cells of larger organisms spend less energy than cells of smaller organisms, but outside their original organisms, they all expend the same amount of energy. In this sense, the cellular hypothesis loses strength in favour of the scaling hypothesis. 
Other empirical examples in this direction can be found in \cite{West2002a,Ahluwalia2017} and in Figure~(\ref{Fig_west}). However, there are studies that support the cellular hypothesis and argue that cells \textit{in vitro} lose their allometry  because they are not performing their normal activities \cite{Glazier2014a}.

Numerous studies have tried to explain the allometric law. Douglas S. Glazier  \cite{Glazier2014a,Glazier2005b,Glazier2018} classified four groups of theories to explain it, which are based on:

\begin{itemize}
	
	\item  \textit{Body surface area}: The ratio between the total body area and the mass of an organism would be the major determinant of its metabolic rate because heat dissipation depends strongly on the body surface area. The mathematical derivations of this theory lead to $\beta =2/3$; and in fact, some groups of experimental data are compatible with this result. More details will be presented in section~(\ref{sec_calor});

\item \textit{Nutrient transport}: Allometric law would be a consequence of the type of the transport network that carries oxygen and nutrients to each cell of the body. Natural selection has made the circulatory system of organisms as efficient as possible, with the result that blood vessels decrease its diameter in a hierarchical, fractal way to the lowest level (capillaries). This idea was proposed by 
Geoffrey West, James Brown, and Brian Enquist  (WBE)\cite{West1997}, who predicted the theoretical value of $3/4$ for the allometric exponent. The premises and results of this theory will be seen in detail in section~(\ref{secao_rede_distribuicao});

\item \textit{System composition}: The theories of this group consider experimental findings that reveal different allometric exponents for isolated organs. This has been observed in the brain, heart, liver, kidneys, spleen, etc., and the behaviour of the exponent of each organ is quite varied\cite{Kestner1936, Darveau2002a, Glazier2014a};

\item \textit{Resource demand}: Allometric law would be a consequence of the energy demand of the cells. This demand would decrease with the size of the organism. This idea is based on the observation that the energy consumption per  \textit{ in vivo} cell decreases with body mass  ($\propto M^{-1/4}$), while the energy consumption in \textit{in vitro} cells does not depend on the mass of the original organism. This absence of allometry \textit{in vitro} would be because the cell is not performing its proper routine activities (it would be in a quiescent state \cite{Glazier2014a}), so it uses the minimum amount of energy necessary for its survival.

\end{itemize}



Finally, what we can say about the allometric exponent is that the value  $2/3$ is compatible with the theories that are based on heat dissipation, while the exponent of $3/4$ is compatible with theories based on nutrient supply networks.
 In this article, we will explore some theories to minimally explore the possibility of explaining the scaling exponent quantitatively.

There are other empirical power laws, and consequently other exponents, that relate certain biological variables, say $y$, with the mass of the organism. These laws have the form 
 $y\sim M^{\alpha}$, where  $\alpha$  is a scaling exponent, which in certain cases is numerically identical to the $\beta$  exponent of the metabolic rate, but which in other cases assumes values that are multiples of $1/4$. 
 That is the case  when the variables in question are related to the \textit{respiratory} or \textit{circulatory systems}. 
 For example, heart rate is related to body mass through an exponent of $\alpha = -1/4$, the time of blood circulation through an exponent of $1/4$, etc. The fact that these exponents are multiples of $1/4$ suggests that metabolism is the master determinant of the other biological variables. In this sense, a change in the metabolic exponent would lead to a systemic change in the other exponents in a kind of  \textit{cascade effect}, as suggested in  \cite{Darveau2002a}. A non-exhaustive list of variables related to the  \textit{circulatory system} and the  \textit{respiratory system} are presented in the tables~(\ref{tabela_sistema_circulatorio}) and~(\ref{tabela_sistema_repiratorio}). They show the empirical and theoretical values of the exponent $\alpha$.


\begin {table*}
\begin{center}
	\begin{tabular}{|c|c|c|}
		\hline   
		$y$ (Circulatory system) &    $\alpha$ (predicted by WBE theory) &  $\alpha$ (empirical) \\
		\hline 
		aortic radius  ($r_0$)  & $3/8 =0.375$  & 0.36 \\
		\hline
		 aortic pressure  & $0 $& $0.032$ \\
		\hline
	 blood velocity in the aorta  ($u_0$) &  $0$  & $ 0.07 $\\
		\hline 
	total blood volume 	 ($V_b$)  & $ 1 $ & $1.00$  \\
		\hline 
	circulation time	 &  $1/4 = 0.25$ & $0.25$ \\
		\hline
	circulation distance ($\sum_{k=0}^K l_k$) & $1/4 = 0.25$ & not available
	 \\
		\hline
	cardiac injection fraction	  & $1$ & $1.03$ \\
		\hline
	 heart rate  & $ -1/4 = -0.25 $&$ -0.25$ \\
		\hline
	 cardiac output	 &$ 3/4 =0.75$   &$ 0.74 $\\
		\hline
	 number of capillaries	($N_c$) & $3/4 = 0.75 $ &  not available
	  \\
		\hline
	capillary density	 &   $-1/12 = -0.083$ & $-0.095$ \\
		\hline
	 oxygen affinity in the blood	 & $-1/12 = -0.083$ &   $-0.089$  \\
		\hline
	service volume radius &   $1/12 = 0.083$ & not available
	 \\
		\hline 
	Krogh cylinder radius  & $1/8 = 0.125$  & not available
	 \\
		\hline
	peripheral resistance	 &  $-3/4 = -0.75$ & $-0.76$  \\
		\hline
	Womersley number & $1/4=0.25$  & $0.25$ \\
		\hline
	 metabolic rate	 & $3/4=0.75$  & $0.74$ \\
		\hline
	\end{tabular}
\caption{ \label{tabela_sistema_circulatorio} Quantities related to the circulatory system and their respective values (theoretical and empirical) of the  $\alpha$ scaling exponent. The theoretical values were obtained from WBE theory, which will be presented in section~(\ref{secao_rede_distribuicao}). Note that in some cases, this exponent is a multiple of $1/4$. The data in this table were extracted directly from \cite{West1997,Peters1983,K.Schmidt-Nielsen1984}.}
\end{center}
\end{table*}

\begin {table*}
\begin{center}
	\begin{tabular}{|c|c|c|}
		\hline   
		$y$ ( respiratory system) &    $\alpha$ (predicted by WBE theory) &  $\alpha$ (empirical) \\
		\hline 
	lung volume	 & $1$ & $1.05$ \\
		\hline
	respiratory rate	  & $-1/4 = -0.25 $ & $-0.26$ \\
		\hline
	volume flow to lung	 & $3/4=0.75$  & $0.80$ \\
		\hline
	interpleural pressure	 & $0$  & $0.004$ \\ 
		\hline
	tracheal diameter	  & $3/8 = 0.375$ & $0.39$ \\
		\hline
	air velocity in the trachea	 & $0$ & $0.02$ \\
		\hline
	 tidal volume	& $1$  &$ 1.041$ \\
		\hline
	dissipated energy	 & $3/4=0.75$  &$ 0.78$ \\
		\hline
	number of alveoli	 & $3/4=0.75$  & not available
	 \\
		\hline
	 alveolar radius	 & $1/12 = 0.083$  & $0.13$ \\
		\hline
	surface area of alveoli	 & $1/6 = 0.1666...$   & not available
	 \\
		\hline
	surface area of the lung	 &  $11/12 = 0.92$ & $0.95$ \\
		\hline
	oxygen diffusing capacity	 &  $1$ & $0.99$ \\
		\hline 
	 total airway resistance	 &  $-3/4 = -0.75$ & $-0.70$ \\
		\hline
	rate of oxygen consumption	 &  $3/4 = 0.75$ & $0.76$ \\
		\hline
	\end{tabular}

\caption{ \label{tabela_sistema_repiratorio} 	
Quantities related to the respiratory system and its respective values (theoretical and empirical)  of the $\alpha$ scaling exponent. As in the circulatory system, in some cases this exponent is a multiple of $1/4$. The theoretical values were obtained from WBE theory, and the data in this table were extracted directly from \cite{West1997,Peters1983,K.Schmidt-Nielsen1984}.} 
\end{center}
\end{table*}



\section{Heat dissipation model}\label{sec_calor}

To stay alive, every organism must convert energy from nutrients into another form of energy, which will be used in vital functions. According to the laws of thermodynamics, processes such as this that convert energy from one form to another must necessarily release heat. For an organism to stay alive, it must eliminate/release this heat at the same rate at which it processes metabolic energy.
In this sense, the heat released by an animal can be understood as a substrate of the energy transformation. 
In fact, one of the ways to quantify the metabolic rate of an endothermic organism is to measure its heat release rate \cite{Lighton2008}.

To begin to understand the consequences of this process of heat production through energy transformation, we will consider two animals of very different sizes, the mouse and the elephant. While a mouse has a mass of the order of $2$g, the elephant has a mass of the order of  $2,000,000$g (two tons); that is, two species with a difference of six orders of magnitude in mass. In a first approximation,
 we could consider that the elephant  expends $10^6$ times much energy than the mouse.    However, this reasoning has the consequence that the amount of heat generated by the elephant would also be $10^6$ times bigger than that of the mouse.
   Therefore, the elephant needs to eliminate all excess heat so that it does not become too hot (which would lead to its death). In fact, the elephant has a large contact surface with the external world (much larger than the mouse), which allows it to dissipate much of the heat produced. Then, the question we need to answer is: Would the size of this surface be sufficient to dissipate the heat generated in the production of metabolic energy?

\begin{figure}
	\begin{center}
\includegraphics[width=\columnwidth]{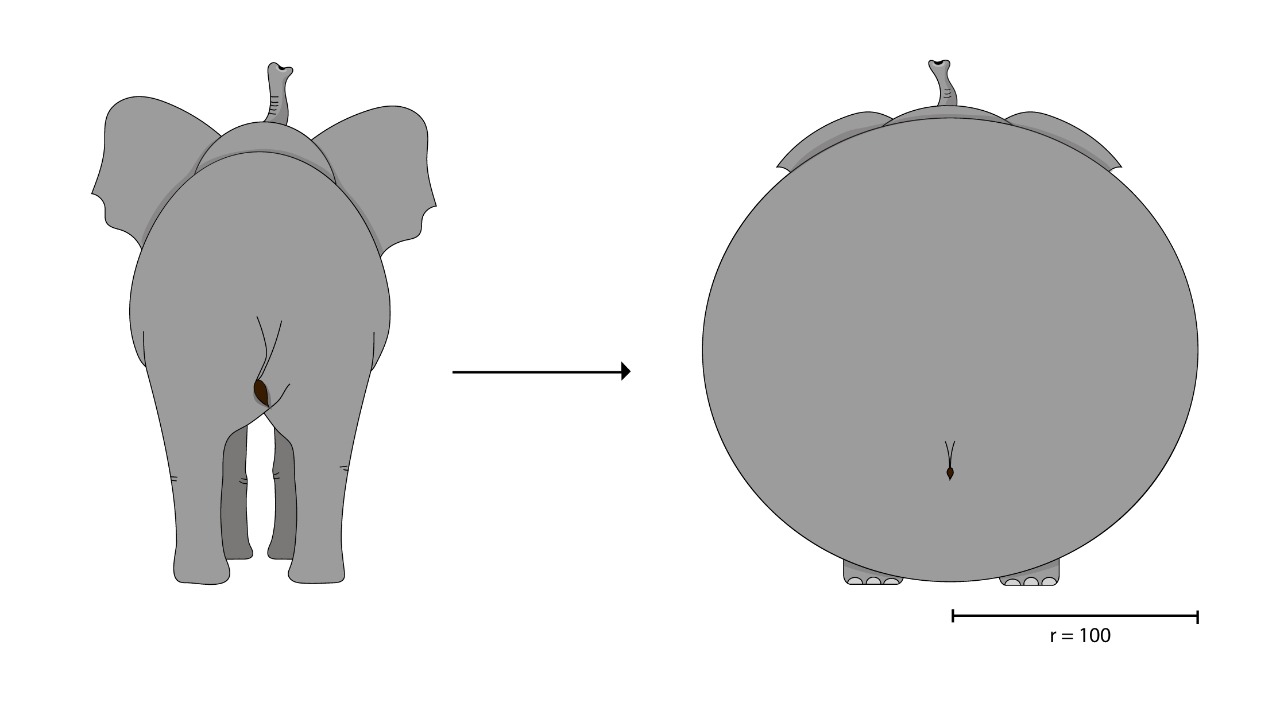}
	\end{center}
	\caption{ \label{fig_rato_elefante}		
		 Rough approximation of the elephant as a sphere. This approximation will allow us to estimate the volume ($\frac{4}{3} \pi r^3$) and the area of the surface ($4\pi r^2$) of this  animal.}
\end{figure}

To try to answer this question, we will consider a very simple model for the contact surface of these animals. Let us consider (roughly) that both the elephant and the mouse  are almost spherical, as in the diagram of Figure~(\ref{fig_rato_elefante}).
 With this approximation, it is much easier to estimate the volume ($V$)  and the area of contact with the external environment ($A$) without loss of generality. These quantities are related to a single linear metric, the radius $r$, through
 $V \approx \frac{4}{3} \pi r^3 \sim r^3$ and  
 $A \approx 4 \pi r^2 \sim r^2$.
 These values are important for determining other properties of the animal. For example, from the volume we can estimate the mass. After all, the mass $M$ of the animal must be proportional to its volume, which implies  $M \sim r^3$.  
Suppose, for convenience, that the mouse has radius $r=1$, given in any unit. Thus, for the elephant volume to be  $10^6$ times greater than that of the mouse, the elephant radius must be approximately $r=100$. Of course, these values are speculative, but we are only interested in the order of magnitude of these numbers; the lack of detail and precision should not compromise the qualitative analysis we are interested in. With this information, we can create Table~(\ref{tabela_rato_elefante})  with the values related to mouses and elephant.

The mass of the animal must be proportional to the volume, so mass of the elephant, as well as its volume, must be $10^6$ times greater than the mass of the mouse. However, the spherical surface area of the elephant is ``only'' $10^4$  times the spherical surface area of the mouse (and not $10^6$ times, as is the case of volume). That is, the surface area increases with $r$ much more slowly than the volume does, which implies that larger animals have relatively smaller surfaces. Quantitatively, we can verify, from the relationships 
 $A \sim r^2$, $V \sim r^3$ and  $M \sim V$, that

 \begin{equation}
 A \sim M^{\frac{2}{3}},
 \end{equation}
which means that the surface area is related to the mass of the animal in a \textit{sublinear} way. In absolute terms, this means that larger animals have greater contact surface area, but these animals have smaller contact area per unit of mass compared to smaller animals.

This result leads us to conclude that the heat dissipation model does not hold. The elephant generates $10^6$ times as much heat as the mouse but radiates this heat on a contact surface of only $10^4$ times that of the mouse. Thus, if the heat dissipation hypothesis were correct, the elephant would be fully carbonized because it has a much smaller contact surface than that required to dissipate all of the heat it produces \cite{mitchell}.

Note that the reasoning in this section is not fully valid because we are considering that the metabolic rate of animals is directly proportional to body mass, which is not true. We will see this in more detail below by using the model proposed by Rubner.

\begin {table}
\begin{center}
	\begin{tabular}{|c|c|c|c|}
		\hline   
		& mouses & elephants \\
		\hline 
		$r$ & $1$ & $100$ \\
		\hline 
		$A$ & $\sim 1$ & $\sim 10.000$ \\
		\hline 
		$V$ & $\sim 1$ & $\sim 1.000.000$ \\
		\hline
		$M$ & $\sim 1$ & $\sim 1.000.000$ \\
		\hline 
	\end{tabular}
	\caption{ \label{tabela_rato_elefante} 
		Values relating to mouses and elephants in a any unit.}
\end{center}
\end{table}

\subsection*{The Rubner model}

\begin{figure}
 	\begin{center}
 	\includegraphics[width=\columnwidth]{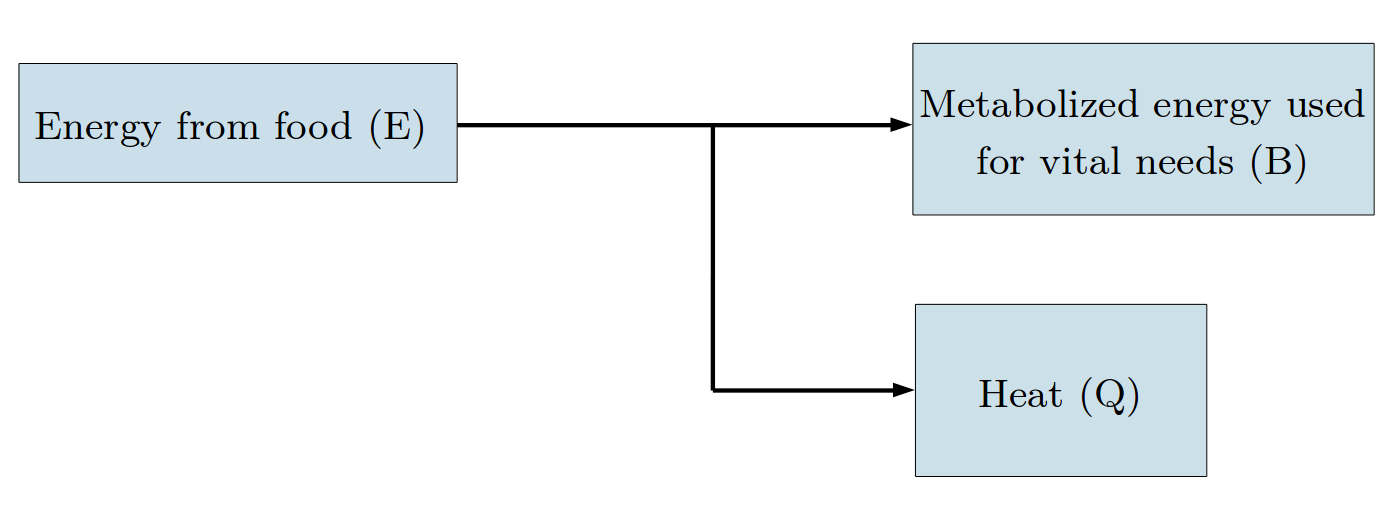}
 	\end{center}
 	\caption{ \label{fig_esquema} 
 		Scheme of energy transformation in organisms.}
 \end{figure}

At the end of the 19th century, Max Rubner postulated that living organisms evolved, by natural selection, to a state in which body mass should follow a surface scaling law and thus be able to radiate excess heat. To understand Rubner’s idea, consider the schematic graph of Figure~(\ref{fig_esquema}), which follows the principle of energy conservation and the second law of thermodynamics. Within the metabolic process, energy from food  ($E$), or any other source, is transformed into:  \textit{useful energy}  ($B$), that is, the energy that will be used for the vital needs of the organism;  and \textit{ heat}  ($Q$), which must be somehow dissipated by the organism. Based on the   \textit{conservation of energy principle},
 $E=B+Q$.

 Rubner hypothetically considered that these three quantities ($E$, $B$ and $Q$) scale with the mass in a similar way and obey the relationships

 \begin{equation}
 E \sim M^{\beta}, 
 \end{equation}
 
 \begin{equation}
 B \sim M^{\beta},
 \end{equation}
 e
 
 \begin{equation}
 Q \sim M^{\beta},
 \end{equation}
 in which $\beta$  is the allometric exponent. In the model considered in the previous section, we hypothesized that metabolic energy was proportional to the mass of the organism, i.e.   $B\sim M$. That is, using the idea expressed in that section, we were considering  $\beta =1$.

To avoid the problem of carbonizing larger animals, the heat must be properly dissipated. Let us consider that the heat  $Q$ is composed of two parts: the dissipated heat $Q_{diss}$ and the heat retained by the organism $Q_{ret}$, and the conservation relationship is valid: $Q = Q_{diss} + Q_{ret}$. In addition, the dissipated heat must be directly proportional to the contact surface of the organism, i.e. $Q_{diss} \sim A$.  We will then look at the ratio $ Q_{diss}/Q$, which will serve as a parameter to measure the efficiency of the organism in dissipating heat. This ratio can give rise to extreme particular cases:

\begin{displaymath}
\frac{Q_{diss}}{Q} = \left\{ \begin{array}{ll}
1 & \Rightarrow   \textrm{all heat produced is dissipated;}  \\
0 & \Rightarrow  \textrm{all heat produced is retained (overheating).}
\end{array} \right.
\end{displaymath}
As $A\sim M^{\frac{2}{3}}$ and, by hypothesis, $Q \sim M^{\beta}$, then  

\begin{equation}
\frac{Q_{diss}}{Q} = M^{\frac{2}{3} - \beta}. 
\end{equation}

\begin{figure}
	\begin{center}
		\includegraphics[width=\columnwidth]{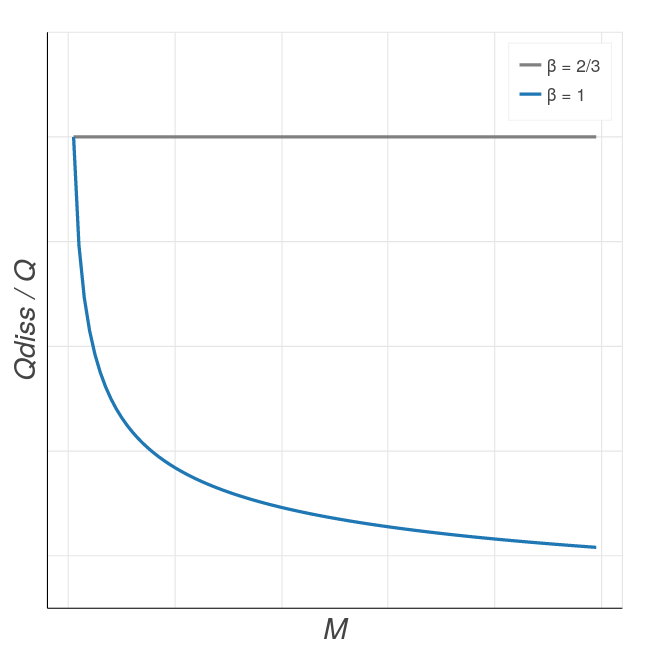}
	\end{center}
	\caption{ \label{Fig_Qdiss}
		Graph of the ratio   $Q_{diss} / Q$  as a function of mass. For $\beta=1$, the ratio and consequently the dissipated heat tend to zero for large  $M$, which in practice would mean overheating in larger animals. In turn, if 
		 $\beta = 2/3$,  then $Q_{diss} / Q$ is independent of the mass of the organism (horizontal line, representing $Q_{diss} / Q \sim M^0$ ). In this case, the animal  should not suffer from overheating if it has a very large mass, as it can proportionally dissipate the same amount of heat as the smaller animals. }
\end{figure}

In the model proposed in the previous section, in which $\beta=1$, we have  $\frac{Q_{diss}}{Q} \sim M^{-\frac{1}{3}}$, which means that the dissipated heat tends to zero for large $M$ (see graph~(\ref{Fig_Qdiss})). This would cause overheating in larger animals, as already discussed. However, as Rubner proposed, if  $\beta = \frac{2}{3}$
then $\frac{Q_{diss}}{Q} \sim M^0 = 1$;  that is, this ratio no longer depends on mass, the individual should not suffer from overheating if it has a very large mass, as it can proportionally dissipate the same amount of heat as the smaller animals.

This Rubner theory, known as the \textit{surface hypothesis}, seems to be quite coherent and was accepted for 50 years. The only problem with this theory is that the allometric exponent predicted by it  ($\beta = 2/3$) is not consistent with much of the experimental data, as verified by Max Kleiber in 1930. $\beta_{exp} \approx 0.74 \approx 3/4$ obtained from the set of data analysed by Kleiber differs from the prediction given by the surface hypothesis. In conclusion, although Rubner's theory has coherent considerations, it is not sufficient to describe the complexity of this scaling phenomenon.

We therefore need another model, another theory, that best describes the empirical evidence. A theory that has the merit of quantitatively explaining the exponent $3/4$ will be introduced in the next section. This theory was developed by 
Geoffrey West, James Brown, and  Brian Enquist,
 and originates from the idea of fractal distribution networks of nutrients to the cells that constitute the organism.

\section{Fractal Distribution Network Model} \label{secao_rede_distribuicao}

At the end of the 1990s, the theoretical physicist Geoffrey West and the biologists
James Brown and  Brian Enquist proposed a model based on the idea of an efficient distribution of nutrients inside an organism
to explain the allometric equation~(\ref{eq_halome}). This model, which we call \textit{ WBE theory}, derives a scaling law between metabolic rate and organism mass
 with exponent
 $\beta = 3/4$ \cite{West1997,West1999,West2004}\footnote{ A critical and alternative review of this model can be found in \cite{Savage2008}.}. 
 This theory has the merit of explaining the scaling phenomenon in a quantitative way, starting from simple and reasonable hypotheses. Before presenting this theory in detail, we will briefly introduce the process of blood and oxygen circulation in an organism.

\subsection*{Circulatory System}

The circulatory system carries blood that contains all the material (glucose, nutrients, oxygen, etc.) that each cell needs to perform its vital functions. Blood distribution begins in the \textit{heart}, a pulsating pump that ensures blood flow throughout the body. From out of the heart comes a large-diameter vessel, the \textit{aorta}, which branches out following a hierarchy of decreasing diameter (aorta → artery → arteriolar → capillary), seeking to reach all parts of the body.  
The oxygen contained in erythrocytes (red blood cells) is transferred by diffusion from the arterial capillary, the smallest circulatory unit, to each of the cells. Soon after, the cell returns carbon dioxide to the same red blood cell, which starts its return journey to the heart through the \textit{venous vessels}. The venous capillary gradually increases in diameter (venous capillary → venula → vein) and returns to the heart.

\subsection*{Respiratory System}

From the heart, the blood saturated with carbon dioxide is directed to the lungs. In the lungs, venous blood receives oxygen captured from air inspiration. The erythrocytes again receive oxygen and release carbon dioxide into the lungs. During the process of air inspiration, this same oxygen-filled air travels inside tubes whose diameters gradually decrease
(trachea → bronchi → bronchioles → alveoli). The alveolus functions as a chamber, and the entire surface of each alveolus is surrounded by capillaries. Hence, the oxygen molecules diffuse from the alveolus to the arterial capillary, and the carbon dioxide exits the venous capillary and diffuses to the alveolus. The concentration of carbon dioxide increases, and then expiration occurs.

\subsection*{Hypotheses of the theory}

Now that some fundamental characteristics of the circulatory and respiratory systems have been presented, we will describe WBE theory. The theory is based on three basic considerations (or hypotheses):

\begin{enumerate}

\item {\bf Fractal distribution network}: The nutrient distribution network, i.e., the circulatory system, has a fractal branching pattern.  
The circulatory system fills the entire volume of the body, carrying nutrients to each of its cells;

\item {\bf  Terminal units (e.g., cells and capillaries) do not vary with the size of the organism}: This hypothesis considers that the quantities related to the last branch of the distribution network – the capillaries – do not vary in relation to the body mass of the individual. These invariant quantities are, for example, the size and mass of a cell and the length, area, and volume of capillaries. Thus, these terminal units function as fundamental building blocks in the construction of any type of biological organism. Some experimental evidence supporting this hypothesis can be found in \cite{Savage2007,Chan2010};

\item {\bf Natural selection and energy minimization}: Natural selection should favour a distribution network that minimizes energy waste (Hamilton principle). An inefficient network for nutrient transport should be eliminated by natural selection.

\end{enumerate}



Before analysing the consequences of these three hypotheses, let us also make some considerations or assumptions:

\begin{itemize}
	
	\item {\bf  Assumption 1}: Total blood volume in an organism is proportional to the mass of the organism;

\item {\bf  Assumption 2}: The metabolic rate is proportional to the blood flow through the aorta of the organism.

\end{itemize}


Assumption 1 comes from empirical evidence (see Table~(\ref{tabela_sistema_circulatorio})). This assumption also arises from the following relationship: Given that the blood volume is proportional to the volume of the organism and that the latter scales linearly with mass, then the blood volume must be proportional to the mass of the organism.

Assumption 2 is based on the idea that blood transports energy (in the form of nutrients) to cells. As all nutrients and oxygen required for metabolic processes are carried by the blood and always pass through the aorta, this assumption occurs naturally.

\subsection{ Modelling the distribution network}

\begin{figure}
	\begin{center}
		\includegraphics[width=\columnwidth]{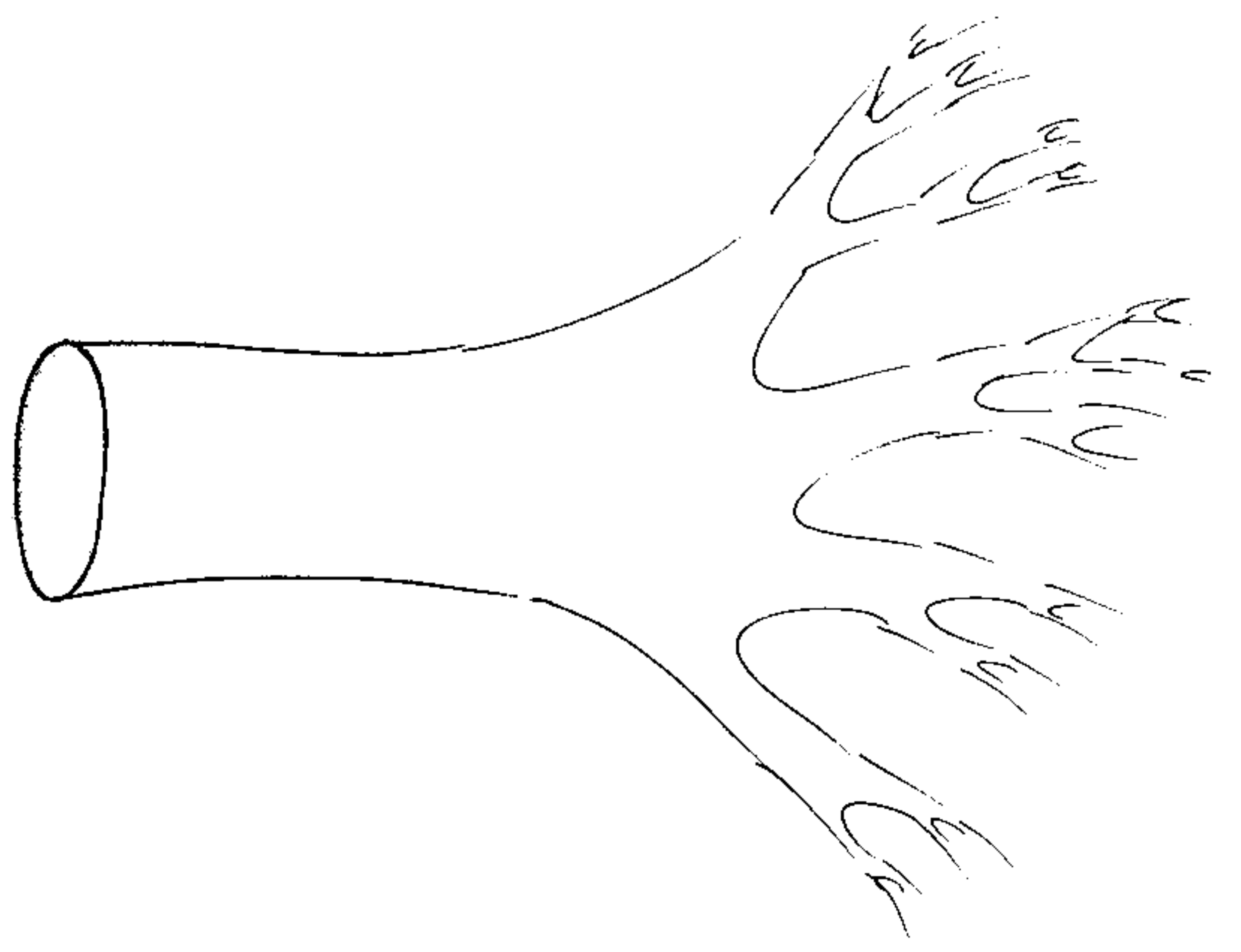}
		\includegraphics[width=\columnwidth]{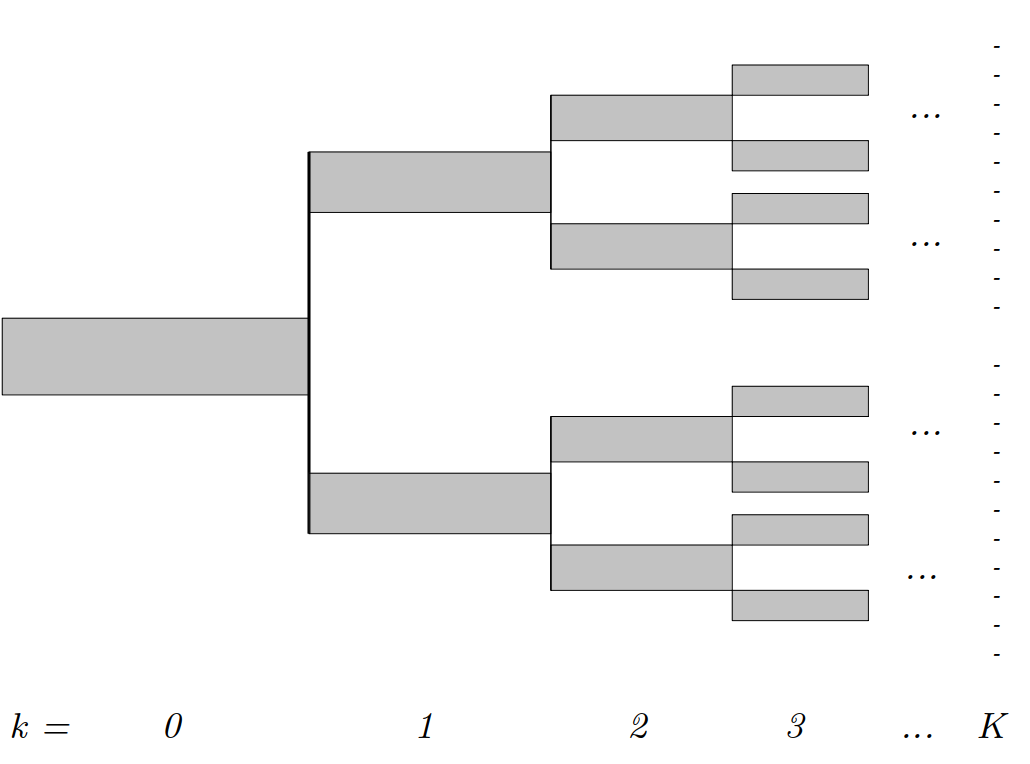}
	\end{center}
	\caption{ \label{fig_rede_fractal} 
		Upper: Fractal form of the branching of the nutrient distribution network (circulatory system). Lower: Very simple model of the fractal form of the circulatory system branching.}
\end{figure}

The nutrient distribution network (circulatory system) presents a fractal form of branching, approximately as described in Figure~(\ref{fig_rede_fractal}) (upper). In a very rudimentary way, we will suppose that these branches can be represented by the stylized model described in Figure~(\ref{fig_rede_fractal}) (lower), where $k$ is an index that represents the level of branching. Note that $k=0$  is the aorta level, while  $k=K$ is the capillary level, which implies that the network is formed by $K+1$ branching levels.
 Each of the blood vessels at a given level branches into $n$ smaller vessels. For example, in Figure~(\ref{fig_rede_fractal}) (lower), we have  $n=2$. That is, the number of vessels of a level has, in this particular case, two times the number of vessels of the previous level. For convenience, $n$ will be considered the same at all network levels. Level $k$ has  $N_k$  vessels, and this number can be determined from the previous level by the recurrence formula

\begin{equation}\label{eq_n}
N_{k+1}= n N_k \, .
\end{equation}
This implies, if $N_0=1$, that

\begin{equation}
N_k = n^k 
\end{equation}
holds.

Thus, the \textit{number of capillaries} $N_c$  of this network, i.e., the number of vessels in the $K$-th level, will be

\begin{equation}\label{Eq_Nc_nK}
N_c \equiv N_K = n^K.
\end{equation}

\begin{figure}
	\begin{center}
		\includegraphics[width=\columnwidth]{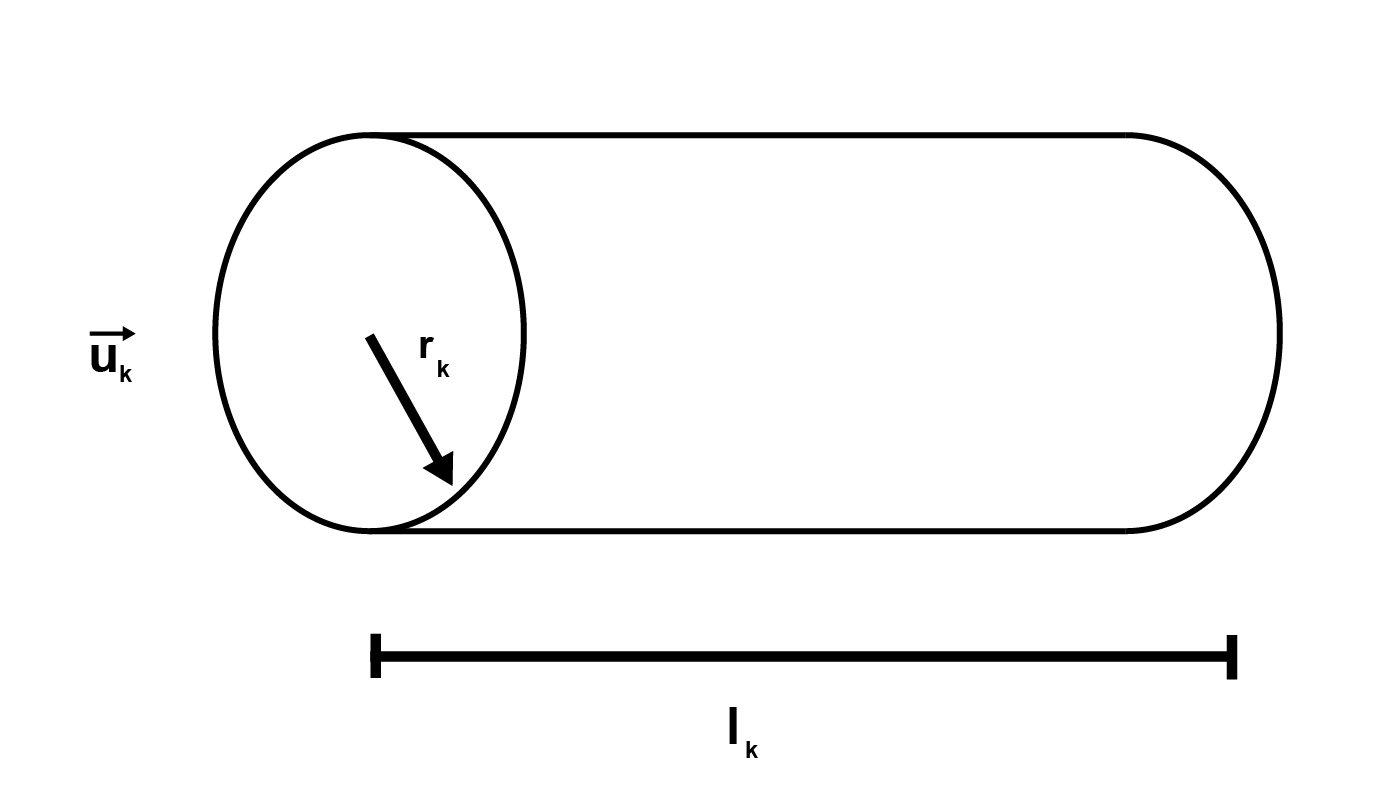}
	\end{center}
	\caption{ \label{fig_cilindro}  
		Cylindrical representation of a blood vessel of the $k$-th level.}
\end{figure}

We will also model the shape of a typical blood vessel of this network. For this, suppose that the vessels have a cylindrical shape, as shown in Figure~(\ref{fig_cilindro}). This figure represents a blood vessel of the $k$-th level, where $l_k$ is the length of the vessel, $r_k$ is its radius, and $u_k$ is the mean velocity of the blood inside that vessel. For example, we can calculate the total distance of blood circulation ($l$) in the organism by the sum of  $l= \sum_{k=0}^K l_k$.

Just as the number of vessels can be written by a recurrence equation, we assume that the length and radius of these vessels also can.
In this sense, one has

\begin{equation}\label{eq_gamma1}
l_{k+1} = \gamma l_k,
\end{equation}
and

\begin{equation}\label{eq_eta}
r_{k+1} = \eta r_k,
\end{equation}
where   $\gamma$ and  $\eta$  are the parameters that relate the subsequent levels. Note that $\gamma$ and  $\eta$  are less than 1 because the vessels of a given level are always smaller than the vessels of the previous level ($l_{k+1}< l_k$ and $r_{k+1}< r_k$).  However, the number of vessels at a given level will always be greater than that at the previous level   ($N_{k+1}> N_k  \Rightarrow  n>1$).

\subsection{Blood flow }

In the appendix~(\ref{apendice_sangue}), we showed that the total volume of blood in the organism, which we call $V_b$, can be written in terms of the parameters introduced above by 

\begin{equation}\label{Eq_Vb}
V_b \propto (\gamma \eta^2)^{-K}. 
\end{equation}
Consider also that $Q_k$ is the blood volume inside of the vessel of level $k$, and then $\dot{Q}_k = \Delta Q_k / \Delta t$ represents the blood flow rate flowing within this tube in a time interval  $\Delta t$. This flow can also be written as

\begin{equation}\label{eq_Qdot}
\dot{Q}_k = \frac{\textrm{blood volume in } k}{\Delta t} = \frac{ (\pi r_k^2)\cdot (u_k \Delta t) }{\Delta t}, 
\end{equation}
where  $\pi r_k^2$ is the cross-sectional area of the vessel and $u_k \Delta t$  is its length ($l_k$).

Eq.~(\ref{eq_Qdot}) leads to 

\begin{equation}
\dot{Q}_k = \pi r_k^2 u_k. 
\end{equation}

As the fluid volume is maintained ($Q_0=N_k Q_k$, for any $k$), the Eq. $\dot{Q}_0 = N_k \dot{Q}_k$ must hold, which can be written in terms of the quantities of the capillary level as 

\begin{equation}\label{Eq_Q_Nc}
\dot{Q}_0 = N_c \dot{Q}_c = N_c \pi r_c^2 u_c.
\end{equation}
Here, $Q_c, \dot{Q}_c,  N_c, l_c, r_c$ and $u_c$ are relative to the capillaries and therefore scaling invariant (by hypothesis 2). 
It is worth adding that experimental observations in mammals suggest that $\dot{Q}_c$  is the same for all species \cite{Savage2008,K.Schmidt-Nielsen1984}.  
We conclude from Eq.~(\ref{Eq_Q_Nc}) that

\begin{equation}\label{Eq_Q0_Nc}
\dot{Q}_0 \propto N_c; 
\end{equation}
that is, hypothesis 2 implies that the blood flow in the distribution network is linearly proportional to the number of capillaries in the organism. Furthermore, by assumption 2, the above result leads us to conclude that the metabolic rate and the number of capillaries scale linearly with each other, that is

\begin{equation}\label{Eq_B2_Nc}
B \propto N_c. 
\end{equation}
From this result, we will derive the numerical value of the allometric exponent in the next section.

\subsection{ Deriving  $\beta=3/4$}

Now we are ready to make a prediction about the allometric exponent $\beta$ from the hypotheses and considerations that make up WBE theory. We have seen by Eq.~(\ref{Eq_B2_Nc}) 
that the metabolic rate is linearly related to the number of capillaries, so given equation~(\ref{Eq_Nc_nK}), we have  $B \sim  n^K$. In addition, by assumption 1, we have  $V_b \sim M$, which implies $n^K \sim M^{\beta} \sim V_b^{\beta}$. Therefore, by result~(\ref{Eq_Vb}), we have

\begin{equation}
n^K \sim (\gamma \eta^2)^{-\beta K}. 
\end{equation}
By taking the logarithm of the two sides of this relationship, we obtain

\begin{equation}\label{Eq_beta}
\beta = - \frac{\ln n}{ \ln (\gamma \eta^2)}. 
\end{equation}
This result tells us that it is enough to know the values of the constants  $n$, $\gamma$ and $\eta$  to calculate $\beta$. That is, the exponent of the allometric law depends only on the constants of the fractal networks that form the circulatory system. This result in itself is a great achievement because it interprets the scaling law in a way completely different from the ideas that permeated the explanations of this phenomenon for more than a century and that were strictly based on the phenomenon of heat dissipation.

\begin{figure}
		\begin{center}
			\includegraphics[width=\columnwidth]{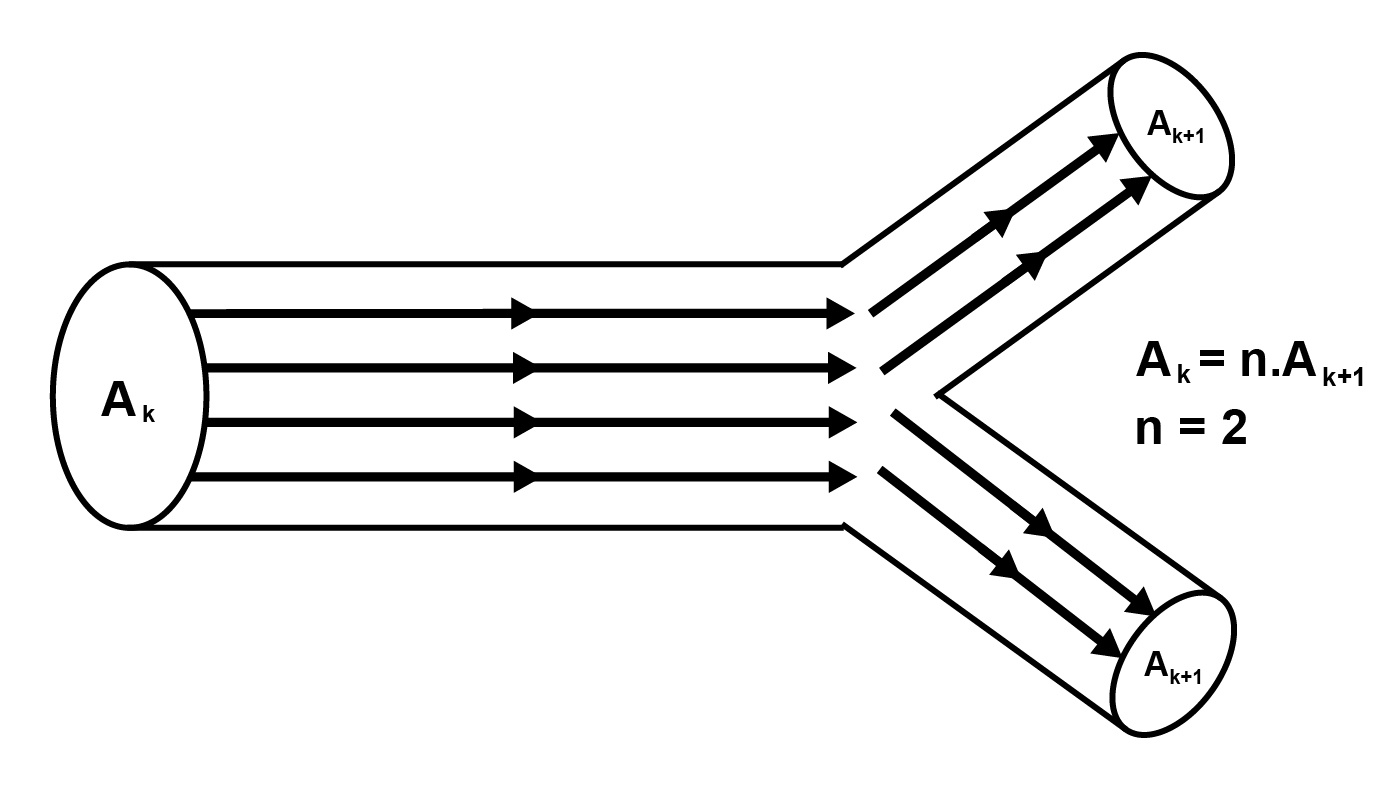}
			\includegraphics[width=\columnwidth]{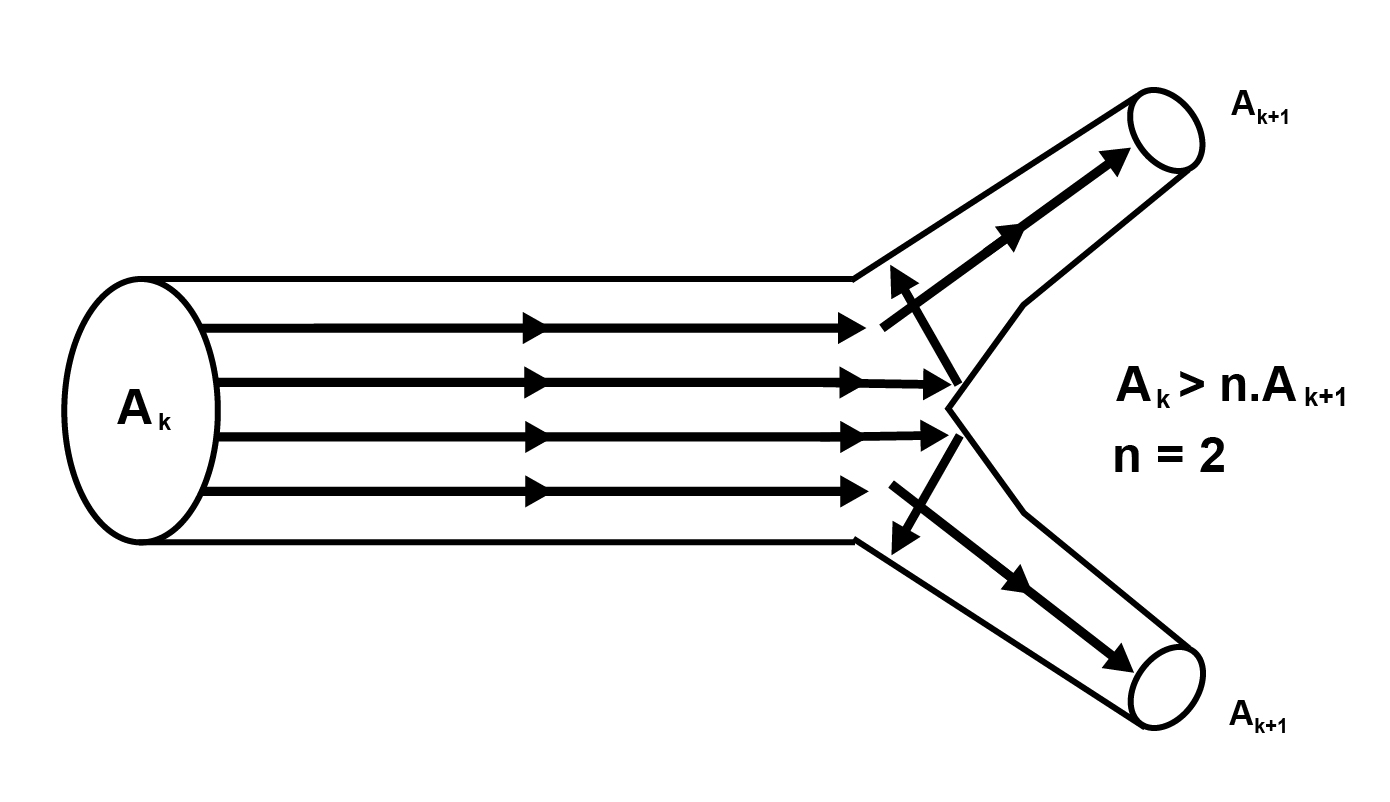}
			\end{center}
		\caption{ \label{fig_conse_area} 
 Scheme of the vessels and the section that passes from one level to another. In the image above, we have an ideal situation in which the cross-sectional area is preserved in the passage from one level to another, i.e., $A_k = n A_{k+1}$. This prevents the blood flow being obstructed. In the image below, in which there is no preservation of the transverse area, the flow is partially interrupted and causes a break in blood propagation. This inefficient type of distribution network should be eliminated by natural selection.}
	\end{figure}

The theory goes even further because we can determine the numerical value of $\beta$ if we consider hypothesis 3 about natural selection, which favours distribution networks that maximizes efficiency and minimizes energy expenditure. In fact, the network that minimizes the loss of nutrients during transport should be the one with the lowest impedance and the one that minimizes the reflection effects of propagation waves. The distribution network that meets these requirements is the one that preserves the transverse area from one vessel level to another, as shown in Figure~(\ref{fig_conse_area}). If there is preservation of the area, then $A_k = n A_{k+1}$ is valid and so

\begin{equation}
\pi r_k^2 = n \pi r_{k+1}^2. 
\end{equation}
Consequently, by inserting $\eta = r_{k+1}/r_k$  (see Eq.~(\ref{eq_eta})) into the above relationship, we obtain

\begin{equation}\label{Eq_eta}
\eta^2 = \frac{1}{n}.
\end{equation}

This result serves to determine how the area, and consequently the radius, of the vessels should vary between the different levels of the vascular network. We will now see how the length of the vessels should vary between these levels. For this, we must understand that the distribution network must be configured to feed/serve each cell of the organism. In this sense, each of the capillaries must feed a set of cells that fill a volume $V_c$, which we will call the \textit{service volume}. This volume should be such that if the organism has  $N_c$ capillaries, then its total volume should be $V= N_c \cdot V_c$. We can repeat this argument for the veins at the level before the capillaries, so that 
 $V= N_{K-1} \cdot V_{K-1}$, where  $V_{K-1}$
 is the service volume of each vein of this level, that is, the volume of capillaries that this vein should serve. This argument can be repeated iteratively for the entire distribution network so that the relationship 
 
 \begin{equation}
 V = N_k V_k = N_{k+1} V_{k+1}
 \end{equation} 
 must be valid for any $k$. By writing the volume  $V_k$ in terms of the length of the vessel at this level $k$ (and by assuming its cylindrical shape as in Figure~(\ref{fig_cilindro})), the above recurrence equation becomes

\begin{equation}
N_k \cdot \frac{4}{3} \pi \left(\frac{l_k}{2}\right)^3 = N_{k+1} \cdot \frac{4}{3} \pi \left(\frac{l_{k+1}}{2}\right)^3 .
\end{equation}
Inputting  $n= N_{k+1}/N_k$ and  $\gamma = l_{k+1}/l_k$   
(Eqs.~(\ref{eq_n}) and~(\ref{eq_gamma1})) into the above equation yields

\begin{equation}\label{eq_gamma}
\gamma = n^{-\frac{1}{3}}.
\end{equation}
Finally, by introducing the results~(\ref{Eq_eta}) and~(\ref{eq_gamma}) into equation~(\ref{Eq_beta}), we conclude that

\begin{equation}
\beta=\frac{3}{4}.
\end{equation}
This result makes the WBE theory one of the most successful theories for explaining scaling laws in biology.
In fact, the result of this theory brought a flavour of exact sciences to biology, in the sense of describing and explaining the phenomena as a deductive consequence of some premises.
In this way, this theory inaugurates an era of systematic understanding of biology,  attempting to find general rules that are valid for a large number of phenomena \cite{Cohen2004,ribeiro_tumor2017, Ribeiro2017}.
Indeed, this is what has happened in Physics for at least 300 years, and the results of this theory show, albeit in a very modest way, that the life phenomenon can also be understood more deeply.

\subsection{Number of capillaries}\label{sec_num_capilares}

We can use this theory to determine how the number of capillaries scales with the organism size. To do this, first consider that the metabolic rate is directly related to blood flow (given assumption 2), so that

\begin{equation}
B \propto  \dot{Q}_0.
\end{equation}
By considering Eq.~(\ref{Eq_Q0_Nc}) and the allometric law ($B \propto M^{3/4}$), we have 

\begin{equation}
N_c \propto M^{\frac{3}{4}}.
\end{equation}
That is, according to the hypothesis and assumptions of this theory, it follows that the number of capillaries must also obey an allometric law with an exponent of $3/4$. However, this prediction has not been verified experimentally. This result leads us to conclude that if this proposed theory is valid, then the intuitive idea that the number of capillaries is linearly proportional to the number of cells (and consequently to the mass) of the organism is wrong.

In addition, the theory predicts an economy of scale because the bigger the animal is,  more cells are fed by one single capillary. 
That is, if $N/N_c$ is the average number of cells fed by one capillary, and $N \sim M$, then

\begin{equation}\label{Eq_num_cells_M}
\frac{N}{N_c} \propto \frac{M}{M^{\frac{3}{4}}} =  M^{\frac{1}{4}}, 
\end{equation}
which shows that $N/N_c$ (cell fed by one capillary) is an increasing function of the organism size.
This is an example of efficiency increasing with size, in a similar way that happens with infrastructure in cities.
In the case of urban phenomena, the bigger the city is, the lesser per-capita infrastructure it demands \cite{bettencourt2007growth, ribeirocity2017,joao_plosone2018}.




\section{Conclusion}\label{sec_conclusion}

	This work intended to give a self-contained insight into the laws governing the relation between some biological properties and organisms' size, particularly the metabolic rate and mass.   We have presented a historical perspective, passing through some data that suggest a power-law behaviour between metabolic rate (and others metrics) and mass, namely allometric law.   We have seen strongly empirical support to this power-law relation, with super-linear behaviour for prokaryotes, linear for protists, and sublinear for vascular organisms. However, the exact numeric value for the scaling exponent is quite uncertain, according to data.

	Some theories to explain the allometric law's sublinear behaviour quantitatively were presented in details in this work.  For instance, the Rubner model, which is based on heat dissipation, comes up with a scaling exponent $\beta =2/3$. 
	It was the most accepted theory to explain the metabolic rate and mass relation for more than 50 years.
	We then presented the WBE theory, based on three primary premisses: i) fractal distribution network; ii) terminal units do not vary with organism size; and iii) natural selection. These premises lead to a scaling exponent $\beta = 3/4$.

	The ideas posted here illustrate science's journey to understand, through a mathematical theory, one aspect of life. Of course, we are still a long way from reaching a level of mathematical description as one has today in physics, for example. However, these theories and ideas gathered here show a giant leap achieved in the last few decades towards a general theory that would explain the phenomenon of life. However, of course, there is still a doubt whether this general theory would, in fact, be achievable, given the great complexity of biology. The next few years will bring us some information about this.

\acknowledgments{
This work was only made possible by the invaluable help of my student Victor Cabral and
Erika Aparecida Costa Lomeu  with the figures presented.  I also want to thank the financial support from the Brazilian agencies CAPES (process number: 88881.119533/2016-01)  and CNPq (process number: 405921/2016-0). 	
}


\appendix

\section{Total Blood Volume in the Organism}  \label{apendice_sangue}

This appendix shows how to determine the total blood volume in the organism, say $ V_b $,  in terms of  WBE theory's parameters.
In fact, it will be shown that 

\begin{equation}
V_b \propto  (\gamma \eta^2 )^{-K}.
\end{equation}

In order to show this relation, consider $V_k$ as the blood volume within a single blood vessel of the level  $k$, which implies  $V_b = \sum_{k=0}^K N_k V_k$. 
As $V_k = \pi r_k^2 l_k$ and $N_k=n^k$  (for  $N_0=1$), one has 

\begin{equation}\label{Eq_Vb2}
V_b = \pi \sum_{k=0}^K n^k r_k^2 l_k. 
\end{equation}

The ratio $ \eta $ between radius of subsequent levels can be written as  $\eta  = r_c / r_{K-1}$, which implies $r_{K-1}= r_c/\eta$,  $r_{K-2}= r_c/\eta^2$, and  so on.
One can then write the recurrence relationship 

\begin{equation}
r_k = \eta^{-(K-k)} r_c.
\end{equation}
Similarly   for the length of the vessels at the level $k$, one has the relation 

\begin{equation}
l_k = \gamma^{-(K-k)} l_c.
\end{equation}
This  way of writing $ r_k $ and $ l_k $ is interesting because they are written in terms of scale-invariant parameters ($ r_c $ and $ l_c $, respectively). Returning to Eq.~(\ref{Eq_Vb2}) one has 

\begin{equation}
V_b = V_c (\eta^{-2K}\gamma^{-K}) \sum_{k=0}^K n^k \eta^{2k} \gamma^k ,
\end{equation}
where $V_c \equiv \pi r_c^2 l_c$ 
is the volume of a capillary and, therefore, scale-invariant. One can then write 

\begin{equation}
V_b \propto  (\gamma \eta^2)^{-K} \sum_{k=0}^K (n \eta^{2} \gamma)^k.
\end{equation}
Here, ``$ = $'' was replaced by ``$\propto $'', giving up the scale-invariant parameters (constants).

The sum in equation above is indeed a geometric progression, with 
initial value  $a_0=1 $, and  common ratio  $ q = n \eta^2 \gamma$. 
Knowing that the sum of a finite geometric progression is  $a_0 (1-q^K+1)/(1-q)$, then

\begin{equation}
V_b \propto  (\gamma \eta^2)^{-K} \left[\frac{1- (n \eta^{2} \gamma)^{K+1}   }{1- n \eta^{2} \gamma} \right].
\end{equation}

Since $ n $, $ \eta $ and $ \gamma $ are constant (by definition) and scale-independent, the denominator term in the above equation can be omitted; that is, we can write simply 

\begin{equation}
V_b \propto  (\gamma \eta^2)^{-K} \left[ 1- (n \eta^{2} \gamma)^{K+1} \right]. 
\end{equation}
Identifying  $N_c = n^K$,  one has $n^{K+1} = n^K n = N_c n$, which leads to  

\begin{equation}
V_b \propto (\gamma \eta^2)^{-K} - N_c n \eta^2 \gamma. 
\end{equation}
Note that the second term to the right of the above equation has only constant or scale-invariant quantities, and therefore one can write

\begin{equation}
V_b \propto (\gamma \eta^2)^{-K},
\end{equation}
demonstrating what was proposed at the beginning of this section. 

One can also calculate the volume of blood in terms of the volume of blood in the aorta. For this, the property 

\begin{equation}
r_k = \eta^k r_0,
\end{equation}

\begin{equation}
l_k = \gamma^k l_0,  
\end{equation}
and
\begin{equation}
V_0 = \pi r_0^2,
\end{equation}
are used, which yield 

\begin{equation}
V_b = \sum_{k=0}^K N_k (\pi r_k^2) l_k = V_0\sum_{k=0}^K (n \eta^2 \gamma)^k.  
\end{equation}
Solving this geometric progression in a similar way to the previous one, one arrives at 

\begin{equation}
V_b = V_0 \left[ \frac{ 1- (n \eta^2 \gamma)^{K+1}  }{ 1- n \eta^2 \gamma  }  \right].
\end{equation}
It is the total blood volume in the organism, written in terms of blood volume in the aorta.


\end{document}